%% file: acl_latex.tex
\pdfoutput=1

\documentclass[11pt]{article}

\usepackage{acl}
\usepackage{graphicx}
\usepackage{amsmath}
\usepackage{booktabs}
\usepackage{comment}

\usepackage{times}
\usepackage{latexsym}
\usepackage{CJKutf8}

\usepackage{todonotes}

\usepackage[T1]{fontenc}

\usepackage[utf8]{inputenc}

\usepackage{microtype}

\newcommand{\coll}[0]{NeuCLIRTech}

\title{\coll: Chinese Monolingual and Cross-Language Information Retrieval Evaluation in a Challenging Domain}

\author{
    \textbf{Dawn Lawrie}$^{\hspace{.1em}\hspace{.1em}{\color{blue}\boldsymbol{\iota}}}$
    \quad
    \textbf{James Mayfield}$^{\hspace{.1em}\color{blue}\boldsymbol{\iota}}$
    \quad
    \textbf{Eugene Yang}$^{\hspace{.1em}\color{blue}\boldsymbol{\iota}}$
    \quad
    \textbf{Andrew Yates}$^{\hspace{.1em}\color{blue}\boldsymbol{\iota}}$
    \vspace{.2em}\\
    \textbf{Sean MacAvaney}$^{\hspace{.1em}\color{blue}\boldsymbol{\gamma}}$
    \quad
    \textbf{Ronak Pradeep}$^{\hspace{.1em}\color{blue}\boldsymbol{\alpha}}$
    \quad
    \textbf{Scott Miller}$^{\hspace{.1em}\color{blue}\boldsymbol{\beta}}$
    \quad
    \textbf{Paul McNamee}$^{\hspace{.1em}\color{blue}\boldsymbol{\iota}}$
    \quad
    \textbf{Luca Soldaini}$^{\hspace{.1em}\color{blue}\boldsymbol{\sigma}}$
    \vspace{.5em}\\
    $^{\color{blue}\iota\hspace{.1em}}$HLTCOE, Johns Hopkins University
    \quad
    $^{\color{blue}\gamma\hspace{.1em}}$University of Glasgow
    \quad
    $^{\color{blue}\alpha\hspace{.1em}}$University of Waterloo
    \vspace{.2em}\\
    $^{\color{blue}\gamma\hspace{.1em}}$Information Sciences Institute, University of Southern California
    \quad
    $^{\color{blue}\gamma\hspace{.1em}}$Allen Institute for AI
    \vspace{.5em}\\
    \texttt{\{lawrie, mayfield, eugene.yang, andrew.yates\}@jhu.edu}
    \\}

\begin{document}

\maketitle

\begin{abstract}

Measuring advances in retrieval requires
test collections with relevance judgments that can faithfully distinguish systems.
This paper presents \coll, an evaluation collection for cross-language retrieval over technical information.
The collection consists of technical documents written natively in Chinese
and those same documents machine translated into English.
It includes 110 queries with relevance judgments.
The collection supports two retrieval scenarios:
monolingual retrieval in Chinese, and 
cross-language retrieval with English as the query language.
\coll{} combines the TREC NeuCLIR track topics of 2023 and 2024.
The 110 queries with 35,962 document judgments
provide strong statistical discriminatory power
when trying to distinguish retrieval approaches.
A fusion baseline of strong neural retrieval systems is included
so that developers of reranking algorithms are not reliant on BM25 as their first stage retriever.
The dataset and artifacts are released on Huggingface Datasets.\footnote{\url{https://huggingface.co/datasets/neuclir/tech}}

\end{abstract}

\input{1_intro}
\input{2_related_work}

\input{3_dataset_creation}

\input{4_results}

\input{5_conc}

\section*{Limitation}
Because relevance judgments are made only for top documents retrieved by dozens of systems at the time of creation,
effectiveness evaluation of systems that were not included as submitted runs to the TREC NeuCLIR track may be underestimated
when they retrieve unjudged relevant documents that are scored as non-relevant.
Researchers using \coll{} should report Judged@20 
as an indicator of possible nDCG@20 underestimation.

\bibliography{custom}

\input{6_appendix}

\end{document}

%% file: 1_intro.tex
\section{Introduction}

This paper presents \coll, a retrieval test collection over technical Chinese documents,
a domain where machine translation currently struggles.
The collection enables evaluation of future first-stage retrieval systems and rerankers
for two retrieval tasks: monolingual Chinese retrieval and cross-language information
retrieval (CLIR) where the queries are in English and the documents in Chinese.
The key distinguishing feature of this task is the technical nature of the documents.
The document collection consists of almost 400,000 abstracts of Chinese academic papers and theses,
drawn from domains such as Chemistry, Physics, and Computer Science.
Figure~\ref{fig:techdoc} shows a sample document from the collection,
along with a machine translation of the document into English.

\coll{} is derived from the work of the TREC Neural Cross-Language Information Retrieval (NeuCLIR) track,
which ran from 2022 to 2024.
\coll{} combines the queries over the two years of that track's technical documents task~\cite{lawrie2023overview, lawrie2024overview}.
The approximately 100 queries with almost 36,000 document judgments provide strong statistical discriminatory power
when trying to distinguish retrieval approaches.
\coll{} provides deep judgments for each query so that
nearly all relevant documents for a query were judged by a human assessor.
It includes a fusion baseline for reranking that combines three modern retrieval approaches:
a single-vector dual-encoder, a multi-vector dual-encoder, and a learned-sparse approach.
Until now, most rerankers have been evaluated using a BM25 first-stage retriever,
exposing powerful rerankers only to documents that have lexical matches with the query.
The main contributions of this work include the combination of query sets into a single evaluation dataset, baselines over more recent CLIR systems for comparison, and 
the fusion baseline, which allows rerankers to be tested on their ability to rank highly relevant 
documents that lack lexical matches with the query.

\begin{figure*}
\footnotesize
    \centering
    \begin{tabular}{l|p{2.5in}|p{2.5in}}
    \toprule
    Key     &   Original  & Google Translation \\
    \midrule
    doc\_id & csl-121374 & csl-121374 \\
    Title & \begin{CJK*}{UTF8}{gbsn}(S)-2-氨基-1,1-二苯基-1-丙醇的合成和外消旋Droxidopa前体的拆分\end{CJK*}
    & (S) -2-amino-1,1-two phenyl-1-propane synthesis and external anti-rotation DroxIDOPA foretophylline is split \\
    Abstract &
    \begin{CJK*}{UTF8}{gbsn}(S)-2-氨基-1,1-二苯基-1-丙醇是一种合成多种手性助剂的重要中间体,也用于外消旋屈昔多巴前体化合物的拆分.从价廉易得的L-丙氨酸出发,通过四步反应制得,总收率55.6\%L-丙氨酸经甲酯化,苄氧羰基保护制得的L-2-苄羰基氨基丙酸甲酯与苯基溴化镁反应制得(S)-2-苄氧羰基氨基-1,1-二苯基-1-丙醇.是在5\%Pd/C催化加氢下脱除苄氧羰基得到标题化合物.该制备方法涉及的中间体及目标化合物易于纯化,总收率高且重现性好.我们用制得的氨基醇能成功地拆分外消旋苏式屈昔多巴前体化合物3-3,4-二苄氧苯基)-N-苄氧羰基丙氨酸.\end{CJK*}
    & \begin{CJK*}{UTF8}{gbsn}(S) -2-amino-1,1-two phenyl-1-propionol is an important intermediate of a synthetic multi-hand assistant. It is also used for demolition Divided. From the price of low-priced L-Alanine, obtained through four-step reaction system, the total revenue is 55.6\%L-alanine through methyl, and the L-2-cymbal base protected by the oxygen oxygenyl The reaction of aminopenate metropolis and phenyl bromide reaction (s) -2-苄 oxygenyl amin amino-1,1-two phenyl-1-propyol. The divisor oxygen cymbal group gets the title compound. The intermediate and target compounds involved in the preparation method are easy to purify, the total yield is high and the reproducibility is good. Daba pre-body compound 3-3,4-two-苄 oxygenyl) - 苄 oxygenyl alanine. \end{CJK*}\\
    Keywords & \begin{CJK*}{UTF8}{gbsn}拆分剂, (S)-2-胺基-1,1-二苯基-1-丙醇, β-氨基醇, 屈昔多巴\end{CJK*}

    & Disassembly, (S) -2-aminel-1,1-two phenyl-1-1-propylene, $\beta$-amino alcohol, Koshidaba \\
    Category & \begin{CJK*}{UTF8}{gbsn}工学\end{CJK*} & Engineering \\
    Discipline & \begin{CJK*}{UTF8}{gbsn}化学/化学工程与技术\end{CJK*} & Chemistry and Chemical Engineering \\
    \bottomrule
    \end{tabular}

    \caption{Example document from the CSL dataset.}
    \label{fig:techdoc}
\end{figure*}

%% file: 2_related_work.tex
\section{Related Work}

Collections for Cranfield-style evaluation of information retrieval effectiveness
have been available for decades~\cite{cranfield,trec1}.
Cross-language collections first appeared in the late 1990s~\cite{DBLP:conf/trec/SchaubleS97}
This retrieval evaluation methodology is well established and widely accepted.

Evaluation collections for retrieval of technical documents are rarer,
yet still available monolingually
(usually with English queries and documents)~\cite{trec-genomics}.
Cross-language evaluation collections for technical documents include the
\citet{10.1145/312624.312730} 
NACSIS test collection. 
This collection has many fewer queries than \coll, with just 21 Japanese queries over approximately 330,000 documents
(in a combination of English and Japanese or either of the languages individually).
The documents are technical papers published by 65 Japanese associations for various fields.

While it tends to focus on high recall settings,
patent retrieval is similar to technical document retrieval.
Like technical documents, patents have a distinct style that is unlike other genres. 
It also uses specialized domain-specific vocabulary.
NTCIR ran several years of a patent retrieval task~\cite{ntcir4-patent,ntcir5-patent,ntcir6-patent}
focusing on search for patent invalidation.
The collection includes Japanese and English documents and queries,
and supports cross-language retrieval.
The CLEF Intellectual Property Track,
which ran from 2009 to 2013,
ran several patent-related tasks,
including search for prior art and passage retrieval
over a collection of English, French, and German patent applications and abstracts~\cite{clefip-2009,clefip-2010,clefip-2011,clefip-2012,clefip-2013}.

%% file: 3_dataset_creation.tex
\section{Dataset Creation}

\subsection{Documents}

The \coll{} documents are abstracts from the Chinese Scientific Literature (CSL) dataset~\cite{li-etal-2022-csl}.
The dataset contains 396,209 journal abstracts from 1,980 academic Chinese journals spanning 67 general disciplines,
where Engineering, Science, Agriculture, and Medicine dominate.
Articles were published between 2010 and 2020.
The abstracts were originally obtained from the National Engineering Research Center for Science and Technology Resources Sharing Service (NSTR).\footnote{https://nstr.escience.net.cn/}

The collection is distributed as JSONL,
a list of JSON objects, one  per line.
Each line represents a document.
Each document JSON structure consists of the following fields:
\begin{description}
\itemsep-0.2em 
    \item [doc\_id:] the assigned id 
    \item[title:]  title of the article
    \item[abstract:] abstract of the article
    \item [keywords:] keywords assigned to article
    \item[category:] category assigned by NSTR
    \item[category\_eng:] human translation of the category
    \item[discipline:] discipline assigned by NSTR
    \item[discipline\_eng:] English human translation of the discipline 
\end{description}

In addition to the document collection itself,
\coll{} includes translations into English of all document fields
obtained from the online \textit{Google Translate} service in  June, 2023.

\subsection{Queries}

Queries were created by 22 graduate students and one postdoc
from the Johns Hopkins University and the University of Maryland, College Park
earning degrees in fields found in the collection.
Creators were hired based on their Chinese language skills and their familiarity with scientific research and paid \$25 per hour for their work.
During an interview, students were asked to describe their research area in both Chinese and English.
They were then asked to choose a research topic they were familiar with,
enter a Chinese language query on that topic
into an interactive search system that returned documents from the CSL dataset,
and read and briefly summarize the top returned documents to determine whether they were relevant to their search.
The purpose of this part of the interview was to ensure that the collection contained documents
related to their area of research
and to verify their ability to understand technical documents written in Chinese.
A Chinese speaker was one of the interviewers.
Of the 22 graduate students,
twelve were Ph.D. and the remaining were Masters students.

Once hired, each annotator participated in a three hour online training session.
During training, they learned how to develop traditional TREC-style information needs
that were scoped at the level of a scientific research paper.
TREC information needs,
usually called \textit{topics},
consist of a short title,
a sentence-length description,
and a paragraph-length narrative.
During ideation, query creators write a description of what they will search for
and a narrative that discusses what is and is not considered relevant to the topic.
During the search phase, the query creator uses a search engine to identify a set of documents to judge.
The creator counts the number of relevant documents in the search results.
Queries with too many relevant documents are revised to be more specific or discarded.
During the revision phase, the description and narrative may be altered
to reflect changes in the way documents were judged.
Finally, the creator writes a title that summarizes the topic in three to five words.
During training, each person worked independently to create their first topic.
Two of the NeuCLIR Organizers reviewed their ongoing work.
This time was used to ensure that the topics had a suitable level of specificity,
and proper tool use.
Including the time spent training, assessors were asked to spend up to a total of ten hours creating five to eight topics.
Three assessors created five topics each,
eleven assessors created six topics,
seven assessors created seven topics,
and two assessors created eight topics,
yielding a total of 146 topics.

An organizer reviewed each English %
title, description, and narrative
to ensure that the topic was sufficiently descriptive,
was grammatically correct,
and contained no spelling errors.
In some cases, assessors were asked to revise topics that appeared to be too vague
or not understandable.
After any revision, assessors verified that the human translation incorporated any changes.
The translations did not undergo any external quality control.

\subsection{Relevance Judgments}

The documents assessors judged for relevance came from the top ranked documents of runs submitted by participating team in the TREC NeuCLIR track of the year in which the query was created.
For queries numbered 1 to 199,
the top twenty documents of each run were assessed for relevance,
while queries numbered 200 and above,
the top thirty-five documents were assessed.
Documents were presented in random order, and assessors were asked to judge all documents for each of the topics they created.

Prior to judging any documents, assessors participated in a second one-hour online training session
on how to assess documents for relevance.
In addition, instructions in Appendix~\ref{sec:rel_instruct} on how to judge relevance were provided.
Relevance grades for the Technical Documents were designed for a searcher interested in scientific abstracts.
Assessors were asked to imagine that they were writing the background section
or the related work section
of a scientific paper on the topic they had selected.
They were asked to evaluate whether they would plan to read the paper being judged based on its abstract
so as to possibly cite the paper in their related work section.

They answered up to two questions about each document
to determine its relevance grade:

\noindent
1. Does this document contain central information?
 \begin{description}
 \itemsep-0.2em 
 \item
 [Yes] Information in the abstract is related to their search topic. 
 \item [No] (0 points)
There is no information in the abstract related to their search topic. 
\item [Unable to judge] (ignored) The document was not viewable in the document viewer panel.
\end{description}
The second question is answered if the first is \textit{yes}:

\noindent
2. How valuable is the most important information in this document?
 \begin{description}
\itemsep-0.2em 
\item
[Very Valuable] (3 points) One would definitely read the paper associated with this abstract
when writing the related work section for this research topic.
\item
[Somewhat Valuable] (1 point) If one had enough time one would read the paper
because it might contain information that could appear in the related work section,
but confidence about that is low.  
\item
[Not that Valuable] (0 points) One is unlikely to read the paper
because one does not expect to find in it information that one would cite in the related work section. 
\end{description}

Assessor progress was tracked;
when very few or an excessive number of relevant documents were found,
or too little time was spent completing the task,
they were asked to rejudge the pool.

Some assessors ran out of time, 
leading to seven unjudged topics.
Eight topics with fewer than three relevant document were removed,
as were sixteen topics for which more than 20\% of the pool was judged to be somewhat or very valuable.
Four other topics were removed because the assessor experienced technical difficulties while completing the task. 

\coll{} includes 110 queries  with relevance judgments.
Each query consists of the concatenation of the topic's title and description.
The collection includes two tab-separated query files for the most common use cases
(one for the original English queries,
the other for human translations of those queries into Chinese),
each line of which contains a query ID and the query text.
A JSONL formatted query file includes many more details,
such as the Google Translations of each field
(to facilitate CLIR experiments on machine translation of the queries).

%% file: 4_results.tex
\input{result_table}

\section{Results}

Table~\ref{tbl:results} summarizes monolingual and cross-language retrieval results.
Qwen3-8B~\cite{qwen3embedding} is the strongest first-stage retriever in both tasks.
For this use case, where the language of the documents comes from scientific abstracts,
MILCO~\cite{nguyen2025milco}, a multilingual LSR model that performs well on the Chinese newswire dataset,
performs below BM25, which tokenizes the text using spaCy~\cite{spacy}.
For consistency with NeuCLIRBench~\cite{neuclirbench},
we use rank-fusion of three systems (Qwen3-8B, PLAID-X, and MILCO) as the first-stage retrieval results for rerankers, despite the fact the fusion performance is lower than Qwen3-8B. 

Not all rerankers can outperform the initial retrieval results.
Particularly in the cross-language task, Jina Reranker,
which is able to maintain roughly the same effectiveness in the monolingual task, becomes worse when crossing languages, highlighting the difficulty of the cross-language task.

%% file: result_table.tex
\begin{table}[t]
\centering
\caption{Retrieval nDCG@20 (nDCG) and Judged@20 (Jdg) on both monolingual and cross-language tasks ordered by CLIR nDCG.}
\label{tbl:results}

\setlength{\tabcolsep}{5pt}
\small
\begin{tabular}{l|cc||cc}
\toprule
{}             & \multicolumn{2}{c||}{Monolingual} & \multicolumn{2}{c}{Cross-Lang.} \\
{}             & nDCG & Jdg & nDCG & Jdg \\
\midrule

BM25 w/ DT            & -- & -- & 0.237 & 1.00 \\
BM25 w/ QT            & 0.290 & 0.99 & 0.274 & 1.00 \\
\midrule
\multicolumn{5}{l}{Bi-Encoders} \\
\midrule
BGE-M3 Sparse         & 0.307 & 0.75 & 0.044 & 0.13 \\
e5 Large              & 0.233 & 0.55 & 0.151 & 0.43 \\
RepLlama              & 0.269 & 0.59 & 0.245 & 0.56 \\
Arctic-Embed Large v2 & 0.359 & 0.79 & 0.262 & 0.67 \\
MILCO                 & 0.253 & 0.56 & 0.264 & 0.66 \\
JinaV3                & 0.382 & 0.79 & 0.305 & 0.68 \\
PLAID-X               & 0.356 & 0.82 & 0.362 & 0.93 \\
Qwen3 0.6B Embed      & 0.402 & 0.81 & 0.377 & 0.77 \\
\textit{Fusion}       & 0.438 & 0.92 & 0.431 & 0.96 \\
Qwen3 4B Embed        & 0.469 & 0.84 & 0.450 & 0.83 \\
Qwen3 8B Embed        & 0.480 & 0.87 & 0.472 & 0.87 \\
\midrule
\multicolumn{5}{l}{Pointwise Rerankers on Fusion} \\
\midrule
Mono-mT5XXL           & 0.456 & 0.94 & 0.407 & 0.93 \\
SEARCHER Reranker     & 0.351 & 0.78 & 0.419 & 0.86 \\
Jina Reranker         & 0.489 & 0.92 & 0.446 & 0.90 \\
Qwen3 0.6B Rerank     & 0.494 & 0.94 & 0.485 & 0.94 \\
Qwen3 8B Rerank       & 0.521 & 0.95 & 0.508 & 0.94 \\
Rank1                 & 0.531 & 0.88 & 0.512 & 0.88 \\
Qwen3 4B Rerank       & 0.522 & 0.94 & 0.526 & 0.94 \\
\midrule
\multicolumn{5}{l}{Listwise Rerankers on Fusion} \\
\midrule
RankZephyr 7B         & 0.434 & 0.92 & 0.432 & 0.96 \\
FIRST Qwen3 8B        & 0.539 & 0.94 & 0.520 & 0.93 \\
RankQwen-32B          & 0.541 & 0.94 & 0.526 & 0.94 \\
Rank-K (QwQ)          & 0.542 & 0.94 & 0.533 & 0.94 \\

\bottomrule
\end{tabular}

\vspace{-0.5em}
\end{table}

%% file: 5_conc.tex
\section{Conclusion}

\coll{} is to our knowledge the first modern cross-language information retrieval evaluation dataset
that focuses on technical documents. %

While the baseline system ordering on \coll{} is similar to that of NeuCLIRBench
(the newswire benchmark variant),
\coll{} poses an additional challenge over CLIR on newswire
because a cross-language technical task is out-of-domain for many current approaches.
The fact that MICLO, a recently developed LSR model, still suffers from low performance on \coll{}
demonstrates that this technical-domain and cross-language challenge requires further investigation.

%% file: 6_appendix.tex
\appendix
\section*{Appendix}
\section{Model Descriptions}\label{sec:models-desc}

\subsection{Sparse Retrieval}

Sparse retrieval methods perform lexical matches between the query and the document. Based on the lexical matches, a
score is produced for each retrieved document. We conduct experiments with BM25 using the Patapsco framework~\cite{costello22-patapsco}. 
Patapsco is a framework built on top of Pyserini~\cite{pyserini} specifically for experiments in 
many languages.
Given the reliance on lexical matches, the input query and documents must be in the same language. 
Some experimental settings such as tokenenization and token normalization are language specific,
while others are language agnostic.
The language agnostic settings include the BM25 hyperparameter settings
where $k1=0.9$ and $b=0.4$ following standard conventions. 
The RM3 query expansion technique~\cite{rm3} 
added ten words based on the top ten documents,
weighting terms from the original query and the expansion terms equally.

For the monolingual retrieval task, all characters were normalized. SpaCy~\cite{spacy} version 0.0.31 was used to perform tokenization. For Chinese,
no stemming was applied.
For the cross-language task, machine translation is used to cross the language barrier. BM25-DT indicates that the documents are translated
into English and the spaCy tokenizer and stemmer are used to process the queries and documents,
which have stopwords removed.
BM25-QT uses machine translation to translate the English query into Chinese and then the Chinese mononlingual setting is followed.

\subsubsection{Multi-Dense Vector Retrieval}

We use PLAID-X~\cite{DBLP:conf/ecir/YangLMOM24,yang2024distillation},
a cross-language variant of the ColBERT model.
PLAID-X is fine-tuned from an XLM-RoBERTa Large model
with cross-language distillation from an mT5 reranker (described below)
using MS-MARCO v1 training queries and passages.
Each document is separated into passages using a sliding window of size 180 tokens with a stride of 90.
At inference time, document scores are aggregated from the passage scores using MaxP~\cite{maxp}.

\subsubsection{Learned Sparse Retrieval}

Learned Sparse Retrieval (LSR) methods represent queries and documents as sparse vectors
where each dimension is associated with a term in a vocabulary.
Some LSR methods produce vectors containing only terms from the input text,
while others use a Masked Language Modeling head to also perform term expansion~\cite{nguyen2023unified}.
To efficiently use these representations for first-stage retrieval,
they can be stored in an inverted index.
We conduct experiments with several LSR methods using Anserini~\cite{yang2017anserini},
which is built on Lucene.

\begin{itemize}
    \item{BGE-M3-Sparse}~\cite{bge-m3} is an LSR method trained as part of the multi-granularity M3-Embedding model. It is based on an XLM-RoBERTa backbone with 0.6B parameters. This method does not perform expansion, so it can only be used for monolingual retrieval.
    \item{MILCO}~\cite{nguyen2025milco} is a multilingual LSR method that performs both query expansion and document expansion. MILCO performs Chinese retrieval by producing sparse vectors that are tied to an English vocabulary, so the model implicitly translates the non-English input. It is based on an XLM-RoBERTa backbone with 0.6B parameters.
\end{itemize}

\subsubsection{Dense Retrieval}
Dense Retrieval (DR) methods represent queries and documents as a single dense vector
in a latent space~\cite{lin2022pretrained}.
To efficiently use these representations for first-stage retrieval,
they can be stored in a vector index that supports approximate nearest neighbor search,
such as FAISS~\cite{douze2025faiss}.
We use fast scan product quantization with FAISS where the number of codes is set to half the embedding dimensionality
and codes are 4 bits each.

\begin{itemize}
    \item Arctic Embed v2~\cite{yu2025arcticembed} is a multilingual model based on a XLM-RoBERTa backbone with 0.6B parameters. We use the \texttt{snowflake-arctic-embed-l-v2.0} checkpoint, which is the largest available Arctic Embed model.
    \item E5~\cite{wang2024multilingual} is a multilingual model based on an XLM-RoBERTa backbone with 0.6B parameters. We use the \texttt{multilingual-e5-large-instruct} checkpoint.
    \item Jina v3~\cite{jinav3} is a multilingual model based on an XLM-RoBERTa backbone with 0.6B parameters.
    \item Qwen3-Embedding~\cite{qwen3embedding} is a family of embedding models trained with synthetic data on top of Qwen3 backbones~\cite{yang2025qwen3}. We use all three model sizes: 0.6B, 4B, and 8B.
    \item RepLlama~\cite{rankllama} is a dense retrieval model based on LLaMA-2-7B. We use the \texttt{repllama-v1-7b-lora-passage} checkpoint trained on MS MARCO passages.
\end{itemize}

\subsubsection{First-Stage Retrieval Fusion}

We use reciprocal rank fusion (RRF)~\cite{cormack2009reciprocal} to combine the rankings
produced by three strong first-stage retrieval models from the three different families described above:
PLAID-X, MILCO, and Qwen-Embedding 8B.
Following common practice and \citet{cormack2009reciprocal}, we set $k=60$ and do not use per-method weights.

\subsubsection{Rerankers}

We rerank the top-$100$ results from the first-stage retrieval fusion
using a range of pointwise and listwise reranking models.
Pointwise models take a query-document pair as input and output a relevance score.
Listwise models take a query and set of documents as input.
Some listwise models output one relevance score for each document,
whereas others output an ordering of their input documents without producing relevance scores.
For models that output an ordering,
we use the partial sorting algorithm from RankGPT~\cite{sun2023chatgpt} to produce a ranking of all 100 results,
where we rerank 20 documents at a time with a sliding window stride of 10 from the bottom of the top 100 candidate documents.

\begin{itemize}
    \item FIRSTQwen-8B~\cite{zijian2024first} is a listwise reranking model built on top of Qwen3 8B~\cite{yang2025qwen3} and trained with the FIRST objective~\cite{reddy2024first}.
    \item jina-reranker-v3~\cite{wang2025jinarerankerv3lateinteractiondocument} is a reranking model built on top of Qwen3 0.6B that can perform pointwise or listwise reranking. In both configurations, the model outputs one relevance score for each document. We use it in a pointwise configuration.
    \item mT5~\cite{bonifacio2022mmarco} is a pointwise reranking model built on top of a multilingual T5 backbone. We use the \texttt{mt5-13b-mmarco-100k} checkpoint with 13B parameters.
    \item Qwen3-Reranker~\cite{qwen3embedding} is a family of pointwise reranking models built on top of Qwen3. We use all three model sizes: 0.6B, 4B, and 8B.
    \item Rank1~\cite{weller2025rank} is a pointwise reranking model fine-tuned to perform reasoning before reranking. Rank1 is built on top of the Qwen2.5 base model~\cite{qwen2.5}. We use the 32B variant in our experiments. 
    \item Rank-K~\cite{yang2025rank} is a listwise reranking model that performs reasoning before reranking using the QwQ 32B reasoning model~\cite{qwq32b}. 
    \item RankQwen-32B is a zero-shot listwise reranking system that uses Qwen3-32B\footnote{\url{https://huggingface.co/Qwen/Qwen3-32B}} without any training to rerank the top-100 documents. %
    This system \emph{does not} use any reasoning.
    \item RankZephyr-7B~\cite{pradeep2023rankzephyr} is a listwise reranking model built on top of Mistral 7B~\cite{jiang2023mistral}. 
    \item SEARCHER~\cite{searcher2024} is a pointwise reranking model built on top of \texttt{Mistral-Nemo-Base-2407}\footnote{\url{https://huggingface.co/mistralai/Mistral-Nemo-Base-2407}} with 12B parameters. It was trained on Persian, Russian, and Chinese translations of MS MARCO passages~\cite{bajaj2018ms}, keeping the queries in English. Documents in the training batch were a mixture of all three languages.
\end{itemize}

\section{Computation Resources Required}

In order to run the experiments reported in Table~\ref{tbl:results}, we used a combination of several GPU types.
We used AMD MI250X GPUs to obtain first-stage retrieval results with the bi-encoder models, which required approximately 500 GPU hours.
We used NVIDIA H100 GPUs to obtain results with the reranking models, which took approximately 150 GPU hours.

\section{Licensing and Intended Use}

Since documents are derived from the CSL collection, we use the Apache 2.0 license, which is the same as the original license.\footnote{\url{https://github.com/ydli-ai/CSL}}
Queries and relevance judgments are under CC-BY, allowing commercial uses with proper attribution.

\section{Instructions for Assessors}

\subsection{Query Development Instructions}
The NeuCLIR track is working on Cross-Language Information Retrieval (CLIR, get it?).  In CLIR, the queries are in a different language than the documents.  To support this research, you will be composing search topics for academic questions in your field study.

A search topic is a statement of a need for information.  It is different from a query. A query is what is typed into the search box; in contrast, the topic is the need that the user is trying to articulate to the search engine.  If I'm looking for an e-bike so I can learn what kinds of e-bikes there are to buy and how much they cost, I might type the query "e-bike models prices".  The topic is what I want to find out about.  The query is a short version of my topic that I think the search engine can work with.

For this task, we will be searching for information in abstracts of theses and articles written in simplified Chinese.  Each of you is working on the project because of your facility in Chinese and your knowledge of a discipline represented in our documents. Your search topics should be something like what you use Google Scholar to find when writing the background section or a related work section for a paper.

You will explore the document collection to find interesting documents that inspire a search topic related to your field of study.  You'll then scope that topic so that it doesn't have too many or too few relevant documents in the collection.  Since you are being asked to create multiple topics, each topic should have a distinct focus; thus your imagined research project would be different for each topic. Then you'll compose a topic statement in English that describes your search need.  Finally, you will translate that description into Chinese.

Each of you will be responsible for creating 5-7 topics over the course of ten hours by 26 June. If you create 7 topics in fewer than 10 hours, please save your remaining hours until August so that you will have more time to do the second task.

The goal of this process is to create a diverse set of search topics, each with a reasonable number of relevant documents.

\subsubsection{Topic development process}

\begin{enumerate}

\item
To start, brainstorm a few ideas for research topics.  Think about possible research projects that you and your colleagues are working on and what papers you might write about them. Your description will describe the information that you would want to include in the related work/background section of this hypothetical research paper. These documents were \textit{written between 2010 and 2020}, so they will not include the most recent information.
\item
On the SEARCH tab enter queries in English and/or Chinese to explore the collection to see if there is anything about the topic.  If you find something you can move to the next step.  You might have to change your topic idea a few times to find a good one.
\item
On the DESCRIBE tab compose the \textbf{description}: a one-sentence statement or question about what you are searching for.  For example, "What government agencies were affected by the WannaCry ransomware attack?" or "I am looking for studies on the detector design of the Beijing Electron-Positron Collider II (BEPCII)."
\item
Write a first draft of the \textbf{narrative} paragraph.  That paragraph should give more details, such as the kinds of information you would accept or would not accept as relevant, what details about the topic you are interested in finding, or anything you think is out of bounds for your topic.  You will revise this in a later step, so don't worry if it's not perfect.  It can even be a comma-separated list at this point.
\begin{enumerate}
\item
Example narrative: The search should return information on detector studies of the Beijing Electron-Positron Collider II (BEPCII). If the result discusses the design or hardware components of BEPCII, but does not include information on detecting, it would be considered as less relevant. If the result includes a specific detector design, but does not mention the electron positron collider, it would be considered irrelevant. Additionally, if the result mainly discusses the software component of the BEPCII, it would be considered as irrelevant.
\end{enumerate}
\item
Lastly, give your topic a \textbf{title}, which should be two to five words describing the topic succinctly, like you might type into a web search engine.  For example, "Beijing electron positron collider detector" might be your title for the last topic in number 3 above.
\item
Return to the SEARCH tab, compose a single query for your topic in Chinese and English or just Chinese.  It might be a translation of your title or it might be something else; it's up to you.  Examine the 30 of the documents retrieved and judge them based on the value of the most valuable information in the document, as follows. First, find the information in the abstract that is closest to the topic you are searching for. Then answer these questions:
\begin{itemize}
\item
Question 1: Does this document contain central information?
\begin{itemize}
\item
\textbf{Yes}: There is information in the abstract related to your search topic
\item
\textbf{No}: There is no information in the abstract related to your search topic.
\item
\textbf{Unable to judge}: the document has not viewable in the document viewer panel
\end{itemize}
\item 
Question 2: How valuable is the most important information in this document?
\begin{itemize}
\item
\textbf{Very Valuable}: you would definitely read the paper associated with this abstract when writing the related work/background section for this research topic.
\item
\textbf{Somewhat Valuable}: If you had enough time you would read the paper, because it might have something that could appear in your related work/background section, but your confidence about that is low.  
\item
\textbf{Not that Valuable}: You are unlikely to read the paper because you don't expect to find information that you would cite in your related work/background section. 
\end{itemize}
\end{itemize}
\item
If the number of very valuable and somewhat valuable abstracts is either less than 1, or greater than 20, the topic needs to be revised.  Broaden or narrow the \textbf{description} sentence if possible and go back to step (6). If you don't see a way to revise, press the "Skip Task" button at the top of the interface to begin a new topic. 
\item
Otherwise, go back and revise your \textbf{narrative} to reflect your final topic.  Make sure to mention edge cases you eliminated by narrowing the topic!
\item
Update the \textbf{title} if necessary. 
\item
Go to the FINISH tab and translate the title and description into Chinese. Create a translation that is a natural expression of the idea in Chinese. Write the translation in Simplified Chinese.
\end{enumerate}

Developing a topic takes about an hour, depending on how long the revision process takes. We would like you to develop 5-7 topics that are neither too narrow or too broad in 10 hours.

\subsubsection{Tool Screenshots}

\begin{figure*}[ht]
    \centering
    \includegraphics[width=\linewidth]{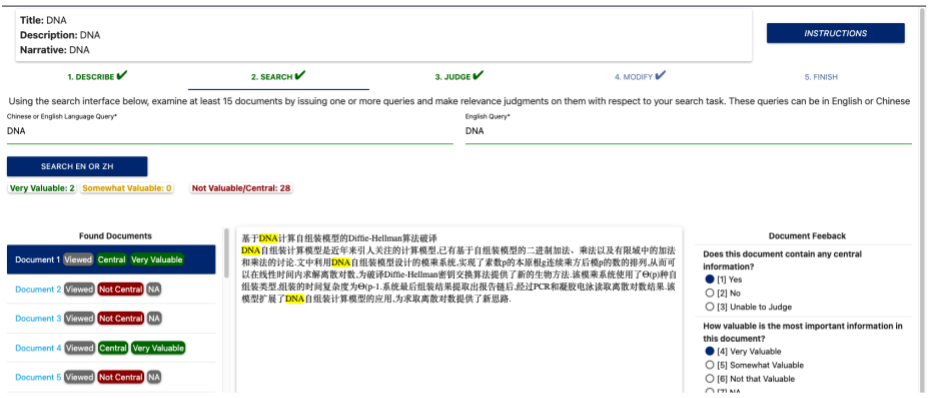}
    \caption{Screenshot of the interface that searches the collection. A query in English or Chinese can be entered. Each ranked document is clicked on the link on the left to display the contents in the middle panel. A document is judged by answering the questions in the right panel.}
    \label{fig:search}
\end{figure*}

\begin{figure*}[ht]
    \centering
    \includegraphics[width=\linewidth]{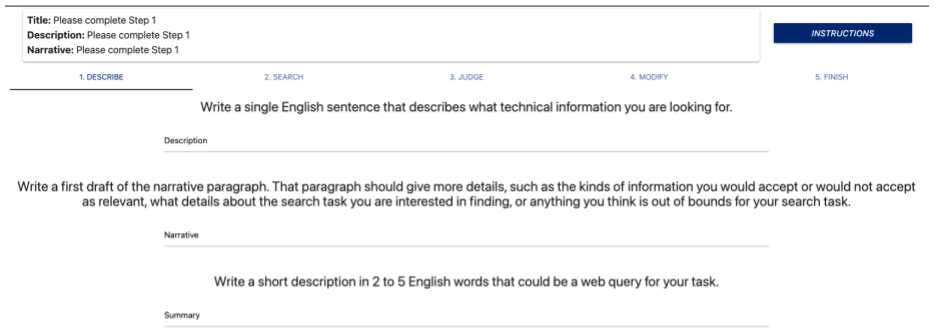}
    \caption{Screenshot of the interface where the assessor records information about the topic.}
    \label{fig:describe}
\end{figure*}

\begin{figure*}[ht]
    \centering
    \includegraphics[width=\linewidth]{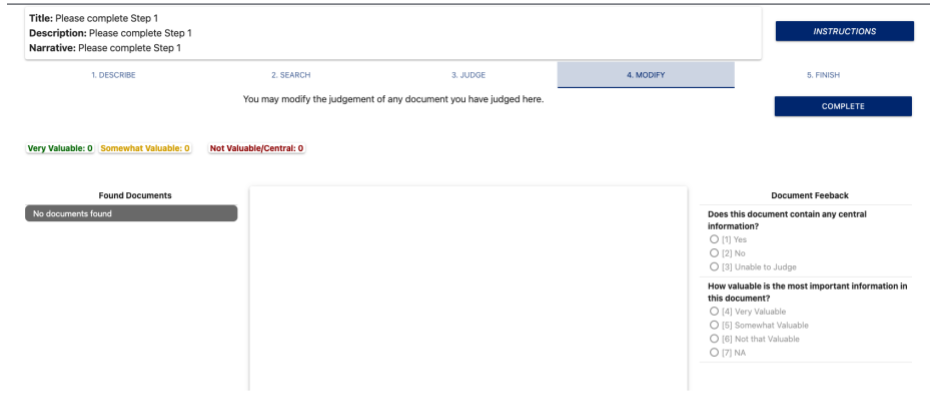}
    \caption{Screenshot of the interface where the assessor can modify the judgments of documents that have already been judged.}
    \label{fig:modify}
\end{figure*}

\begin{figure*}[ht]
    \centering
    \includegraphics[width=\linewidth]{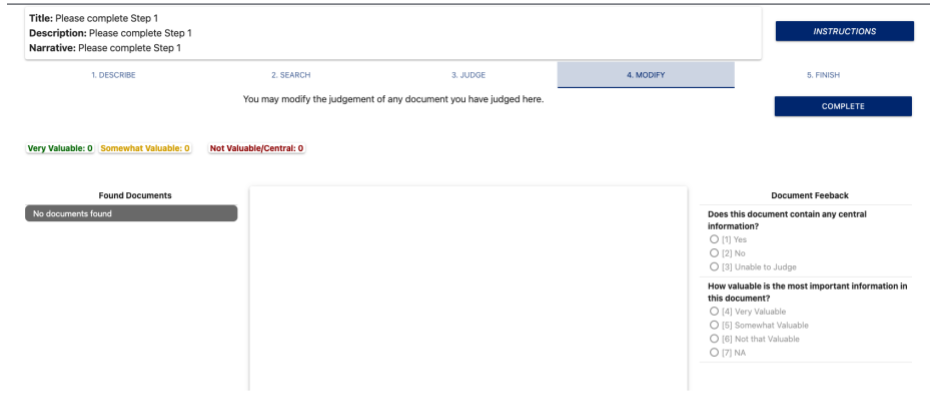}
    \caption{The screenshot of the final tab where assessors complete the task by entering translations of the title and description in Chinese. They had to rate how easy it was to find central information and optionally could add a comment about the topic.}
    \label{fig:finish}
\end{figure*}

After brainstorming to determine a potential topic is Step 1, the assessor used the search tab shown in Figure~\ref{fig:search} to search the collection of documents for information that is relevant to their topic idea. Then
the assessor moved on the describe tab shown in Figure~\ref{fig:describe} and completed Steps 3 to 5 before returning to the search tab to judge 30 documents.
The counter shown in Figure~\ref{fig:search} tracked their progress on the task. If they ever needed to revise their judgments on the documents, the modify tab in Figure~\ref{fig:modify} could be used. The task was completed on the Finish tab shown in Figure~\ref{fig:finish} where translations were entered.

\subsection{Relevance Assessment Instructions}
\label{sec:rel_instruct}

\subsubsection{Overview}
The NeuCLIR task is concerned with Cross-Language Information Retrieval (CLIR):  using search queries in one language to search for documents in another language.  In our case, the searches are in English, and the documents are in Chinese. What makes this task unique is that the documents are abstracts of articles and theses, and therefore contain more different vocabulary than newswire.

You may remember that about a month ago you developed a set of \textbf{search topics}.  Each topic represents an “information need,” a gap in knowledge leading us to search.  A topic is more than just a query – it defines what you were looking for and what kinds of information you would find relevant.

In this second part of the task, you will be assessing documents for relevance.  For each topic, you will receive a \textbf{pool} of documents that were retrieved by TREC participant systems.  You will judge each document as to whether it is relevant to your topic.

You will judge documents for the topics you created.

\subsubsection{Tour}
\begin{figure*}[h]
    \centering
    \includegraphics[width=\linewidth]{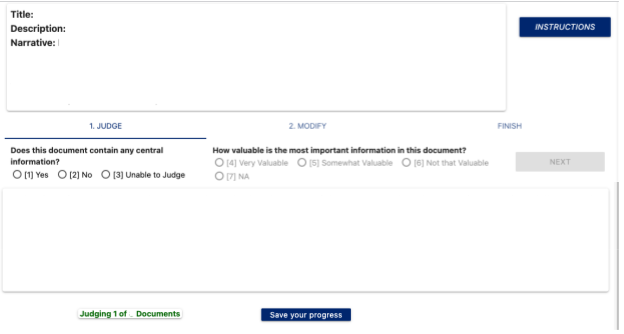}
    \caption{The screenshot of the judge tab where assessors record relevance judgments for all the top ranked documents requiring assessment. The title, description, and narrative for the topic was always visible while judging documents. Below the relevance questions, the document was displayed. Once a document was judged, the next button would become active. A counter at the bottom tracked their progress.}
    \label{fig:judge}
\end{figure*}

After you accept a task, you should see Figure~\ref{fig:judge} at the top of your browser.

On the first screen you will see the English description of the topic, the two questions that you must answer about the document, the contents of the document, a counter, and a button to “Save your progress.” Once you answer the questions, the “NEXT” button will turn blue so that you can go to the next document. 

You begin on the JUDGE tab. This is where you will see each document in turn. The MODIFY tab is used to review or change the judgment for a document. Once you are done judging the pool, you will be brought to the FINISH tab. If you want, you may leave a comment. Then click SUBMIT. Only submitted tasks will be recorded. If you click on the FINISH tab before you judge the entire pool, you will have to confirm that you want to submit the task early. Please do not do this without first consulting with Dawn Lawrie.

The numerals next to the buttons can be used to select the radio buttons in lieu of a mouse click. For example, if I wanted to judge something “Very Valuable,” I could type 1 followed by 4. In this way, your mouse can stay on the next button if the documents aren’t too long.

\subsubsection{Relevance}
What does it mean for a document to be relevant to your topic?  Imagine that you are writing the background section or the related work section for a scientific paper. You will use the best documents in the pool as citations for your paper.

You will answer two questions about each document.
\begin{itemize}
    \item 
Question 1: Does this document contain central information?
\begin{itemize}
    \item 
\textbf{Yes}: There is information in the abstract related to your search topic
\item 
\textbf{No}: There is no information in the abstract related to your search topic.
\item 
\textbf{Unable to judge}: the document has not viewable in the document viewer panel
\end{itemize}
\item
Question 2: How valuable is the most important information in this document?
\begin{itemize}
    \item 
    \textbf{Very Valuable}: you would definitely read the paper associated with this abstract when writing the related work/background section for this research topic.
\item
\textbf{Somewhat Valuable}: If you had enough time you would read the paper, because it might have something that could appear in your related work/background section, but your confidence about that is low.  
\item 
\textbf{Not that Valuable}: You are unlikely to read the paper because you don’t expect to find information that you would cite in your related work/background section. 
\end{itemize}
\end{itemize}

\subsubsection{Tips}
It is best to complete the judgments for the entire topic without closing your browser. However, there is a “Save your Progress” button. This saves your work to the cookies in your current browser; it won’t be accessible from other browsers or if you clear your cookies. Thus, it will usually save your work, but is not robust. 

Judge each document \textbf{independently}.  It must be valuable (or not) on its own.  That also means that if you see valuable information repeated, it’s still valuable and should be judged the same.  

How to judge each document is completely up to you, as the user of the search system and author of the imagined paper.  However, it’s critical that you try to be \textbf{consistent}.  If you judge something to be very valuable once, apply that same standard to other documents.  If some other piece of information is not relevant or not that valuable, judge it the same way the next time you see it.

It’s important to make steady progress through your pool.  If you’re not sure how to judge a document, use your best judgment. You may write down the number of the document so that you can revisit it on the modify tab later. 

The vast majority of documents you see will not be relevant.  Aside from those, many documents may be topically related, but not be very valuable.  Typically there are only a few somewhat valuable or very-valuable documents. However, recall that when creating the topic you found between 1 and 20 documents about your topic. These documents are in the pool and we expect that they would continue to be valuable, so when I say a few, it could be a few dozen documents.

You will find that documents that are obviously not relevant take very little time to identify.  Make sure you at least skim to the end of each document to make sure there isn’t anything topical buried in there.  If you want to look for specific terms, use Find in your browser.

Once you reach ten hours of working on these tasks, finish the one you are on and stop. 

\section{Data Contents}

The data we collected from assessors was converted into scores of 0, 1 or 3 for each document. The data is distributed as a file in the standard qrels format with a judgment per line. This judgment consists of a "$<$topic id$>$ 0 $<$document id$>$ $<$score$>$.